# A Possible Biogenic Origin for Hydrogen Peroxide on Mars: The Viking Results Reinterpreted

Joop M. Houtkooper[1] and Dirk Schulze-Makuch[2]

[1]Center for Psychobiology and Behavioral Medicine
Justus-Liebig-University of Giessen, Otto-Behaghel-Strasse 10F,
D-35394 Germany
E-mail: joophoutkooper@gmail.com

[2]School of Earth and Environmental Sciences
Washington State University
Pullman, WA 99164, USA
Email: dirksm@wsu.edu

**The adaptability of extremophiles on Earth raises the question of what strategies putative life might have used to adapt to the present conditions on Mars. Here, we hypothesize that organisms might utilize a water-hydrogen peroxide ($H_2O$-$H_2O_2$) mixture rather than water as an intracellular liquid. This adaptation would have the particular advantages in the Martian environment of providing a low freezing point, a source of oxygen, and hygroscopicity. The findings by the Viking experiments are reinterpreted in the light of this hypothesis. Our conclusion is that the hitherto mysterious oxidant in the Martian soil, which evolves oxygen when humidified, might be $H_2O_2$ of biological origin. This interpretation has consequences for site selection for future missions to search for life on Mars.**

The Viking landers on Mars remain the only direct attempt to detect life on another world. All three experiments conducted by the landers observed chemical changes that indicated the possible presence of life, although the expected signals were not as large as expected for a biological response and tapered off after a while, casting doubt on a biological explanation. This led to a consensus view that Viking had detected reactive, oxidizing surface chemistry, but not biochemical metabolic processes (*1*).

We propose here a reinterpretation of the Viking results, based on the assumption that microorganisms on Mars produce hydrogen peroxide to generate an $H_2O$-$H_2O_2$ intracellular solvent for biochemical processes selected by and adapted to the unique Martian environment.

Current environmental conditions near the surface of Mars are not incompatible with life. Various survival studies exposing terrestrial microbes to simulated near-surface conditions on Mars have revealed remarkably high survival rates below very shallow soil (*2,3*). In regard to oxidant-tolerance, some soil bacteria survive and grow to stationary phase in 30,000 ppm $H_2O_2$ (*4*). McDonald et al. (*5*) tested the stability of organic macromolecules subjected to oxidation stress by 30 % $H_2O_2$ in water at three diffferent temperatures relevant to Martian environmental conditions. Their data suggested that some organic macromolecules are stable against oxidation on the Martian surface, at least in the polar regions, over the entire history of Mars. Mixtures of $H_2O_2$ and $H_2O$ would have remarkably useful properties for any organism in need to adapt to Martian environmental conditions.

Mixtures of $H_2O_2$ and $H_2O$ freeze at temperatures significantly below the freezing point of water. The lower eutectic point lies at -56.5°C for a mixture with 61.2 weight % $H_2O_2$ (6). Mixtures with a high $H_2O_2$ concentration tend to supercool, sometimes resulting in the formation of glasses, down to liquid-air temperatures (7). Thus, putative Martian organisms could stay completely functional at temperatures far below the freezing point of water and even survive lower temperatures as the formation of ice crystals and piercing of cellular membranes would be prevented. $H_2O_2$-$H_2O$ mixtures are slightly acidic, the 60 weight % mixture has a pH of 4.5. $H_2O_2$-$H_2O$ mixtures tend to be hygroscopic because of the lower water vapor partial pressure in equilibrium with the liquid, as compared with water, which would offer the opportunity for an organism to scavenge water molecules from the Martian atmosphere. These considerations point to at least the possibility that organisms might use $H_2O_2$, not only in storage as a convenient source of oxygen, but also as a major component of their intracellular fluid.

An intracellular $H_2O_2$-$H_2O$ mixture would not only provide a source of oxygen and be a favorable adaptation to cold temperatures, but also convey hygroscopic abilities to the putative Martian organisms. At little metabolic cost they would be able to scavenge the atmosphere for the little water vapor present. On the other hand, this ability would mean a vulnerability for exposure to liquid water. They would be susceptible to death by hyperhydration. This could occur when the organisms are exposed to liquid water or even to a relatively warm atmosphere saturated with water vapor. At death, the cellular contents would be set free. This could be in the form of organic compounds, but $O_2$ might well be a component. Furthermore, it is likely that an exothermic reaction between the $H_2O_2$ and the organics would occur, and the organism would transform into $CO_2$, $O_2$, and water vapor (and some nitrogen and minor constituents).

The $H_2O_2$ could be produced biochemically by the organisms themselves, through energy obtained from sunlight. A gross metabolism pathway could follow

$$CO_2 + 3\ H_2O \leftrightarrow CH_2O + 2\ H_2O_2 \tag{1}$$

The equation would proceed to the right using sunlight as an energy source and to the left in darkness or when work is exerted. However, when exchange with the environment is to be avoided, the sugars could be oxidized to formic acid

$$CH_2O + H_2O_2 \leftrightarrow HCOOH + 2\ H_2O \tag{2}$$

However, this would quickly acidify the cell, like lactic acid does in terrestrial animals. In an alternative reaction, $H_2O_2$ could serve as a source of energy by simply decomposing into water and oxygen:

$$2\ H_2O_2 \rightarrow 2\ H_2O + O_2 \tag{3}$$

Based on the freezing point of a $H_2O$-$H_2O_2$ mixture, the organisms would be well-adapted if their intracellular fluid contains a substantial concentration of $H_2O_2$. This is if an adaptation to these high amounts of $H_2O_2$ in the intracellular fluids would not require too high of a cost in terms of energy requirements such as the production of stabilizing compounds. Metabolic activity could theoretically occur at temperatures down to -56°C (217 K), if an eutectic mixture is used. Temperatures lower than the freezing point of the $H_2O_2$-$H_2O$ eutectic could also be withstood because of likely supercooling of the liquid. An upper temperature limit depends on the stabilization mechanism, which must consume energy at a faster rate at high temperatures. As the cell contents are energy-rich, higher temperatures than usually occur on Mars may be withstood for brief periods of time.

While the life detection experiments conducted by the Viking landers were generally interpreted as a failure to detect life based on the biochemistry of microorganisms on Earth, doubts and inconsistencies about those results remain. Particularly,

(1) While no organic molecules were detected by gas chromatography-mass spectrometry (GC-MS), the requisite sensitivity may not have been achieved at the time.
(2) Chemical explanations for the Viking lander experiments (particularly the evolution of $O_2$ upon wetting) require a strong oxidizer at sufficiently high concentration, which has still not been identified.
(3) There is no satisfactory explanation for the 30 % rise in $CO_2$, the near doubling of $N_2$, or the surprising large rise of $O_2$, from 4 nmol to about 520 nmol, in the Gas Exchange Experiment (*8*).
(4) No convincing mechanism had been proposed for the small but significant synthesis of organic material in the Pyrolytic Release Experiment (Table 1). This amount could not come from the synthesis by UV radiation since an optical filter to screen out the UV wavelengths below 320 nanometers was included in the experiment.
(5) The production of gas recorded from the Labeled Release (LR) nutrient when it was placed on Martian soil at both lander sites was significant. Decreases of released gas were observed at secondary injections. The reactant in the Mars soil was completely unreactive at the sterilizing temperature of 160°C. In contrast, exposure to 18°C for two Martian days did not inhibit the reaction.

The failure to detect organic molecules by the GC-MS during the Viking mission to Mars was surprising, especially because some $2.4 \times 10^8$ g of reduced carbon falls on Mars each year via asteroids, comets, and other planetary material (*9*). The common assumption is that all the organic material near the surface is oxidized by $H_2O_2$ and other strong oxidizing compounds. Based on the reactivity of the surface measured by the Viking Gas Exchange experiment (GEx), the amount of $H_2O_2$ on the Martian surface was estimated to be between 1 ppm (*10*) and 250 ppm (*4*). Yet, photochemical processes generate $H_2O_2$ in the atmosphere at a much lower rate in the parts per billion range. Atmospheric $H_2O_2$ abundances vary between 20 and 40 ppb by volume over the planet (*11*), which appears to be a maximum concentration occurring during favorable weather conditions (*12*). Thus, there is a case to be made not only for the missing organics but also for the missing $H_2O_2$.

The biological explanation of the lack of detected organics by GC-MS could be that the oxidizing inventory of the $H_2O_2$-$H_2O$ solvent well exceeded the reducing power of the organic compounds of the organisms. Upon heating, therefore, the putative organisms might have auto-oxidized catastrophically, leaving the gases as detected by the GEx experiment plus very little solid residue without or with only little organic content. The negative result of the GC-MS (*13*) is therefore not a very reliable estimate of an upper bound on the biomass in the soil. A reasonable alternative explanation of the missing organics using a purely chemical explanation was advanced by Benner et al. (*14*), who suggested that any organics on the surface of Mars would undergo a diagenesis to metastable compounds of carboxylic acids derivatives and would not be easily detected by GC-MS. Explanations to the five questions presented above in terms of the $H_2O_2$-$H_2O$ hypothesis and traditional chemical explanations are provided in Table 2.

The fact that $O_2$ evolved from soil samples upon humidification deserves particular scrutiny in order to evaluate if this is in accordance with a biological rather than a chemical origin. The release of $O_2$ was a surprise for the mission scientists, who then became convinced in a chemical origin as terrestrial life is not known for originating $O_2$ upon wetting. Our interpretation, however, is that under a dry or slightly humid atmosphere the release of oxygen would be a consequence of a metabolic pathway of putative $H_2O_2$-$H_2O$ based Martian organisms (Reaction 3), while under wet conditions it would indicate the decomposition of organisms by hyperhydration after being exposed to excess water. A similar inference can be drawn from the Pyrolytic Release (PR) Experiment. Wetting all but inhibited any organic synthesis reaction in the Utopia 2 sample and the following Utopia 3 sample (Table 1), while organic synthesis reactions did occur under dry conditions (Chryse 1, and possibly to some minor degree also for Chryse 3 and 4, and Utopia 1). The mission scientists had problems to explain this phenomenon. Horowitz et al. (*15*) called it "startling", while Klein (*16*) felt that a explanation for this phenomenon "remains obscure".

Not all results can be satisfyingly explained with our hypothesis. For example, the differences in amplitude of the response of the PR experiment remain a puzzle. However, this might be understood by assuming that the Martian surface is not covered by a homogeneous population of organisms. And some chemical reactions certainly play a role in the response to the Viking lander experiments. In view of the $H_2O_2$-$H_2O$ hypothesis on Martian life, the Viking experiments were both too warm and too wet. Especially the combination of high temperatures (relative to average Martian conditions) and saturation with water vapor is an extremely unmartian condition and both the GEx and the LR experiment employed this condition. The interpretation of these two experiments might well not be related to heterotrophic metabolism (as the tests were designed for), but in terms of coping and failing to adverse conditions. The putative Martian organisms were overwhelmed by too much water vapor, a condition against which they had no defense, so that they failed because of too high of an osmotic pressure.

If we assume $H_2O_2$-$H_2O$ based life as a working hypothesis, we conclude that adding water would only be of limited benefit to growth. At the test cell temperatures of about $10^oC$, organisms might survive 50% humidity for some time, whereas 100% humidity at that temperature seems to be fatal within a few days at most. The hygroscopicity of the $H_2O_2$-$H_2O$ mixture and the probable lack of a mechanism to exclude too much water are the likely causes of the sensitivity for water, even as vapor. The argument that Viking detected reactive chemistry rather than biology based on the fact that there is no known Earth organisms that can be shown to reproduce all Viking results is wanting. Likewise, is the argument that Viking discovered biology, because there is no known mineral or reactive Earth-analog chemistry that produces all Viking results. Any explanation of the Viking results must be intrinsically linked to the Martian environment with its differing geochemistry and also organisms, if they exist.

$H_2O_2$-$H_2O$ solutions are mostly known as desinfectants and sterilizing agents on Earth. Thus, the compatibility of $H_2O_2$ with biological processes might seem questionable. However, some microbial organisms produce hydrogen peroxide (e.g., certain *Streptococcus* and *Lactobacillus sp.* (*17,18*), while other microbes utilize $H_2O_2$ (e.g., *Neisseria sicca*, *Haemophilus segnis*, *H. parainfluenzae*, *Actinomyces viscosus*, and *Staphylococcus epidermidis* (*18*). The microbe *Acetobacter peroxidans* even uses $H_2O_2$ in its metabolism (overall reaction $H_2O_2(aq) + H_2(aq) \leftrightarrow 2H_2O$ (*19*)). However, the high reactivity of $H_2O_2$ poses a problem to most microorganisms, which control it by the use of stabilizing compounds. Colloidal silicate and pyrophosphate often are used in commercial products, compounds such as phenacetin, an aromatic amine (N-ethoxy-acetanilide) may be more applicable to organisms. Most microbes that come into contact with $H_2O_2$ protect themselves with scavenging enzymes such as catalase, glutathione peroxidase, and peroxiredoxin. $H_2O_2$ is commonly used as defense mechanisms by microbes, antibodies (*20*), immune cells, and even certain insects. The Bombardier beetle, *Brachinus crepitans*, for example, has in its posterior a chitinous chamber in which a mix of fluids can be injected, one of which is a 25% solution of $H_2O_2$ (*21*). This is combined with hydroquinone and a catalyst to produce a steam explosion in the chamber, which can be directed at a pursuing predator. The uses of $H_2O_2$ in biology are surprisingly diverse. Mammalian cells are known to produce $H_2O_2$ to mediate diverse physiological responses such as cell proliferation, differentiation, and migration (*22,23*) and biological redox reactions catalyzed by $H_2O_2$ typically involve the oxidation of cysteine residues on proteins (*24*). Thus, high concentrations of $H_2O_2$ can be produced and utilized biochemically even in terrestrial organisms. There does not appear to exist a basic reason why $H_2O_2$ could not be used by biology. On Earth, utilizing $H_2O_2$ in the intracellular fluid has little advantage with regard to temperature and availability of oxygen and water, thus the majority of Earth organisms never developed extensive adaptation mechanisms. On Mars, on the other hand, directional selection may have favored organisms which developed on an early warm and wet Mars to adapt to the progressive cooling and desiccation of Mars.

The utilization of $H_2O_2$ is not without some drawbacks. $H_2O_2$ decomposes spontaneously, thus an organism needs some mechanism to stabilize the $H_2O_2$. The situation is even more demanding for photoautotrophic organisms exposed to sunlight, which on Mars includes a considerable flux of UV with wavelengths down to about 200 nm. $H_2O_2$ will decompose under UV radiation and has to be protected by pigments in the cellular membrane or by an active stabilization mechanism. This does not necessarily require chlorophyll, but could involve bacteriorhodopsin embedded in the cell membrane such as in halophilic organisms, or involve some inorganic compound such as cycloocta sulfur for efficient UV

6protection (*25*). However, to date, no suitable UV protection compound has been identified to exist on Mars, perhaps indicating that any such organisms, if they exist, would have to pursue an endolithic lifestyle comparable to the microbes in the Antarctic Dry Valleys (*26,27*). Notably, these problems are of lesser magnitude at lower temperatures requiring fewer resources.

In contrast to water-based organisms, the putative Martian autotrophs would need to avoid liquid water. However, in the generally arid environment, it is beneficial if the water vapor partial pressure is above about 50 % a significant fraction of time. Also, a generally low ambient temperature is beneficial in view of the stabilization of the cellular contents. If organisms on Mars exist that use the proposed biochemistry, they would likely be active in colder areas on Mars with high water vapor concentrations as would be expected along the polar ice fringes. The Martian tropical areas may be warmer and drier than optimum, thus were not the optimal location for a life detection experiment such as Viking.

Our hypothesis of Martian organisms that would utilize a $H_2O_2$-$H_2O$ mixture as intracellular liquid is of great consequence for future missions searching for extant life on Mars. Rather than exploring in the equatorial belt, where temperatures might allow liquid water to exist for brief periods of time, life may well exist in temperate or sub-arctic regions, where temperatures are colder and the atmosphere contains more water vapor. These concerns would also have to be addressed in future sample return missions to Mars.

**Table 1.** Data from the Pyrolytic Release Experiment (*28*). The Conditions Column indicates whether the lamp was on or off, whether or not water vapor was injected, and whether the soil sample was heat-sterilized (control is 175$^o$C for 3 hours). The radioactivity of Peak 2 column represents organic matter synthesized from the labeled gases.

| Experiment | Conditions | Peak 2 (count/min) |
|---|---|---|
| Chryse 1 | Light, dry, active | 96 +/- 1.15 |
| Chryse 2 | Light, dry, control | 15 +/- 1.29 |
| Chryse 3 | Light,dry, active | 27 +/- 0.98 |
| Chryse 4 | Light,dry, active | 35 +/- 1.6 |
| Utopia 1 | Dark, dry, active | 23 +/- 1.7 |
| Utopia 2 | Light, wet, active | 2.8 +/- 0.92 |
| Utopia 3 | Dark, dry, active | 7.5 +/- 2.5 |

**Table 2.** Explanations for some remaining questions after Viking

| Question | Chemical Explanation | $H_2O_2$-$H_2O$ hypothesis |
|---|---|---|
| Lack of organic molecules | The organics have been oxidized to nonvolatile salts of benzenecarboxylic acids, and perhaps oxalic and acetic acid (*14*). | Upon death of the organisms, the organics spontaneously are oxidized by $H_2O_2$ with no or very little organic residue. Non-biology bound organic molecules are oxidized chemically (*14*) and/or consumed by organisms. |
| Lack of oxidant | There is some yet unidentified mechanism producing $H_2O_2$ or other oxidants. The oxidant might be present in form of a compound that has no analog on Earth. | The $H_2O_2$ in the $H_2O_2$-$H_2O$ mixture is part of the biochemistry of the putative Martian organisms. It would provide sufficient oxidizing potential to explain the Viking results. |
| Release and Partial Resorption of $O_2$, $CO_2$, and $N_2$ in the GEx experiment | Evolution of $O_2$ on humidification was suggested to involve one or more reactive species such as ozonides, superoxides, and peroxides (*8*). $CO_2$ production in the wet mode can be interpreted to be related to the oxidation of nutrient organic compounds (*29*) and $N_2$ release can be interpreted to be related to an initial $N_2$ desorption from soil by water vapor and subsequent resorption in liquid water (*29*). | The release of $O_2$ (and possibly $CO_2$ to a lesser degree) can be interpreted as the result of an energy-producing metabolism. Upon humidification it could point also to the decomposition of dying Martian biota, as could the increase of $N_2$. The decrease of $N_2$ can be understood as biological fixation, a possibility also entertained by Oyama et al. (*29*). |
| Synthesis of organic material in PR experiment | No consistent explanation has been provided, but attempts to explain the observations include instrument malfunction, incorporation of $^{14}CO$ into carbon suboxide polymer preformed on the Martian surface, and reduction of $^{14}CO$ by $H_2O_2$ in the surface material (*15*) | Some of the putative organisms were able to metabolize and synthesize organic compounds before they died being overwhelmed by water. |
| Responses in the Labeled Release experiment | Laboratory test on Earth using inorganic oxidants and clay minerals simulated many of the key findings (*1*). | Limited metabolism (*30*) before the organisms died due to hyperhydration, osmotic pressure, and/or heat shock. |